\documentclass[prd,aps,showpacs,preprintnumbers,amssymb]{revtex4}
\usepackage{axodraw}
\usepackage{color}
\usepackage{epsf}

\def\e3p{$\eta \rightarrow 3 \pi$}

\begin{document}
\title{%
\hfill{\normalsize\vbox{%
\hbox{}
 }}\\
{Exotic MSSM}}
\author{Renata Jora $^{\it \bf a}$~\footnote[1]{Email:
 rjora@theory.nipne.ro}}
 \affiliation{$^ {\bf \it a}$ National Institute of Physics and Nuclear Engineering, PO Box MG-6, Bucharest-Magurele, Romania.}
\date{\today}

\begin{abstract}
We present an exotic version of the MSSM with the same particle content but with some of the states with negative kinetic energies.
We discuss how the model can work quantum mechanically and also resolve some of the pressing issues of the regular MSSM.

\end{abstract}
\maketitle

\section{Introduction}
Recent experimental data from the Large Hadron Collider (LHC) found no evidence \cite{LHC1}-\cite{LHC2} for the supersymmetric extensions of the standard model.
In particular natural SUSY \cite{Papucci} implies light masses for the gluinos, stop, sbottom and Higgsinos whereas the superpartners can be as heavy as we want.
However experiments already excluded the stop and sbottom masses up to about 200 GeV putting a fair amount of pressure on these models. Furthermore the
upper limit that the MSSM poses on the mass of the light Higgs boson requires  the addition of beyond the MSSM physics at least in the Higgs sector, if the
sbottoms and stops are kept light.

Some of these issues can be resolved elegantly without the introduction of new particles if one considers  a simple version of the MSSM in which the sign of the kinetic term of one
of the Higgs multiplets is changed. Of course in order to preserve supersymmetry some of the corresponding interaction terms should be also introduced with opposite signs.

In particular we know that the regular MSSM gives a bad prediction at tree level for the observed 125 GeV Higgs boson:
\begin{eqnarray}
m_H\leq m_Z \cos(\beta).
\label{s4343}
\end{eqnarray}

 A wrong kinetic sign for the second Higgs doublet transforms the $\cos(\beta)$ into a $\cosh(\beta)$ thus predicting the right answer, $m_h=125$ GeV.
In general we know that hyperbolic functions in particle physics are associated with  branes and extra dimensions so this kind of feature might be justified if the model is viewed
as a low energy description of an underlying string-brane theory with reminiscent effects on the second Higgs multiplet.

Moreover we will show in section VII that although the states of negative energies seem peculiar at the level of classical field theory, all the inconsistencies can be eliminated at the level of second quantization. In  particular, if we can accept that field theory with both positive and negative kinetic energies can have stable oscillations \cite{Trodden}
around  a saddle point of the potential (corresponding respectively to positive and negative masses) we can see how at the quantum level the ill defined states can be cured by using
a simple and old paradigm: the negative energies are associated with time inversion and time inversion can be replaced by CP- conjugation if CPT is to be respected in the theory.
We show in detail how this works for our exotic MSSM in sections V and VI.

Section VIII contains some phenomenological aspects and the conclusions.

\section{Short review of a gauge and chiral supermultiplet lagrangian}
We start with the simple case of a gauge supersymmetric theory with a single chiral multiplet.
The Lagrangian has the well known expression:
\begin{eqnarray}
{\cal L}={\cal L}_{gauge}+{\cal L}_{chiral}-\sqrt{2}g(\Phi^*T^a\Psi)\lambda^a-\sqrt{2}g\lambda^{\dagger a}(\Psi^{\dagger}T^a\Phi)+g(\Phi^*T^a\Phi)D^a
\label{lagr54646}
\end{eqnarray}

where,
\begin{eqnarray}
&&{\cal L}_{gauge}=-\frac{1}{4}F^a_{\mu\nu}F^{\mu\nu a}+i\lambda^{\dagger a}\bar{\sigma}^{\mu}\Delta_{\mu}\lambda^a+\frac{1}{2}D^aD^a
\nonumber\\
&&{\cal L}_{chiral}=-\Delta^{\mu}\Phi^*\Delta_{\mu}\Phi+i\Psi^{\dagger}\bar{\sigma}^{\mu}\Delta_{\mu}\Psi+F^*F.
\label{lagr5454}
\end{eqnarray}

Here $\Phi$, $\Psi$ form the chiral multiplet and $A^a_{\mu}$ and $\lambda^a$ form the gauge supermultiplet (We refer the reader to the thorough review in \cite{Martin}, \cite{Aitchinson}
for details.) This Lagrangian is invariant under the following supersymmetry transformations:
\begin{eqnarray}
&&\delta A^a_{\mu}=\frac{1}{\sqrt{2}}(\epsilon^{\dagger}{\bar \sigma}^{\mu}\lambda^a+\lambda^{\dagger a}{\bar \sigma}_{\mu}\epsilon)
\nonumber\\
&&\delta\lambda^a_{\alpha}=\frac{i}{2\sqrt{2}}(\sigma^{\mu}{\bar \sigma}^{\nu}\epsilon)_{\alpha}F^a_{\mu\nu}+\frac{1}{\sqrt{2}}\epsilon_{\alpha}D^a
\nonumber\\
&&\delta D^a=\frac{i}{\sqrt{2}}(-\epsilon^{\dagger}{\bar \sigma}^{\mu}\Delta_{\mu}\lambda^a+\Delta_{\mu}\lambda^{\dagger a}{\bar \sigma}^{\mu}\epsilon),
\label{tr5454}
\end{eqnarray}
and
\begin{eqnarray}
&&\delta\Phi=\epsilon\Psi
\nonumber\\
&&\delta\Psi_{\alpha}=-i(\sigma^{\mu}\epsilon^{\dagger})_{\alpha}\Delta_{\mu}\Phi+\epsilon_{\alpha}F
\nonumber\\
&&\delta F=-i\epsilon^{\dagger}{\bar \sigma}^{\mu}\Delta_{\mu}\Psi+\sqrt{2}g(T^a\Phi)\epsilon^{\dagger}\lambda^{\dagger a}.
\label{res434343}
\end{eqnarray}

The transformations in Eq(\ref{tr5454}) are the same as for a single uncoupled gauge supermultiplet. In comparison the transformations in Eq(\ref{res434343}) are different
from those specific to a single chiral supermultiplet: the partial derivative is changed to a covariant derivative and $\delta F$ receives an additional term. One can conclude that the gauge supersymmetry transformations do not change in the presence of the chiral supermultiplet. The term,
\begin{eqnarray}
{\cal L}_{chiral}-\sqrt{2}g(\Phi^*T^a\Psi)\lambda^a-\sqrt{2}g\lambda^{\dagger a}(\Psi^{\dagger}T^a\Phi)+
g(\Phi^*T^a\Phi)D^a.
\label{tr544454}
\end{eqnarray}

is then shown to be invariant separately under the combined supersymmetric transformations.
We continue with a short review of the Higgs sector of the MSSM Lagrangian.
\section{The Higgs sector of MSSM}

The MSSM contains plenty of chiral multiplets corresponding to the usual fermion and Higgs boson parts. The superpotential is given simply by,
\begin{eqnarray}
W_{MSSM}={\bar u} y_u Q H_u-{\bar d} y_d Q H_d-{\bar e}y_e L H_d+\mu H_u H_d.
\label{superp0990}
\end{eqnarray}
Here $H_u$, $H_d$, Q, L, ${\bar u}$, ${\bar d}$, ${\bar e}$ are the chiral multiplets corresponding to the two Higgses, the left handed quarks and leptons and right handed quarks and leptons respectively.
This superpotential will lead, among other things, to a mass term for higgsinos and Higgses:
\begin{eqnarray}
&&-{\cal L}_{higgsino\, mass}=\mu(\tilde{H}^{+}_u\tilde{H}^-_d-\tilde{H}^0_u\tilde{H}^0_d)+c.c.
\nonumber\\
&&-{\cal L}_{Higgs\,mass}=|\mu|^2(|H^0_u|^2+|H^+_u|^2+|H^0_d|^2+|H_d^-|^2)
\label{massterms545}
\end{eqnarray}

The MSSM scalar potential in the presence of soft supersymmetry breaking terms turns out to be
\begin{eqnarray}
&&V=(|\mu|^2+m_{H_u}^2)(|H_u^0|^2+|H_u^+|^2)+(|\mu|^2+m_{H_d}^2)(|H_d^0|^2+|H_d^-|^2)+b(H_u^+H_d^--H_u^0H_d^0)+c.c.+
\nonumber\\
&&+\frac{1}{8}(g^2+g'^2)(|H_u^0|^2+|H_u^+|^2-|H_d^0|^2-|H_d^-|^2)^2+
\frac{1}{2}g^2|H_u^+H_d^{0*}+H_u^0H_d^{-*}|^2.
\label{pot6565}
\end{eqnarray}

Here the term proportional to $|\mu|^2$ comes from the F term (superpotential) and those proportional to $\frac{g^2+g'^2}{8}$ and $\frac{g^2}{2}$ from the D contribution in the Lagrangian.
The $m_{H_u}^2$, $m_{H_d}^2$ and b terms are soft supersymmetry breaking term.
The minimum conditions read:
\begin{eqnarray}
&&m_{H_u}^2+|\mu|^2-b\frac{v_d}{v_u}-\frac{g^2+g'^2}{4}(v_d^2-v_u^2)=0
\nonumber\\
&&m_{H_d}^2+|\mu|^2-b\frac{v_u}{v_d}+\frac{g^2+g'^2}{4}(v_d^2-v_u^2)=0.
\label{min676}
\end{eqnarray}

The mass matrix for the neutral scalars is computed to be:
\begin{eqnarray}
M^2_{h,H}=\left[
\begin{array}{cc}
b\frac{v_d}{v_u}+\frac{g^2+g'^2}{2}v_u^2&-b-\frac{g^2+g'^2}{2}v_uv_d\\
-b-\frac{g^2+g'^2}{2}v_uv_d&b\frac{v_u}{v_d}+\frac{g^2+g'^2}{2}v_d^2
\end{array}
\right].
\label{sc34343}
\end{eqnarray}

The mass matrix for the pseudoscalars is:
\begin{eqnarray}
M^2_{A^0,G^0}=
\left[
\begin{array}{cc}
b\frac{v_d}{v_u}&b\\
b&b\frac{v_u}{v_d}
\end{array}
\right],
\label{ps546}
\end{eqnarray}

whereas the charged scalars have the mass matrix:
\begin{eqnarray}
M^2_{h_u^+,H_d^-}=
\left[
\begin{array}{cc}
b\frac{v_d}{v_u}+\frac{g^2}{2}v_d^2&b+\frac{1}{2}g^2v_uv_d\\
b+\frac{1}{2}g^2v_uv_d&b\frac{v_u}{v_d}+\frac{g^2}{2}v_u^2
\end{array}
\right].
\label{ch6565}
\end{eqnarray}

From these mass matrices one can easily derive the corresponding masses:
\begin{eqnarray}
&&m^2_{G^0}=m^2_{G^{\pm}}=0
\nonumber\\
&&m^2_{A^0}=\frac{2b}{\sin(2\beta)}=2|\mu|^2+m^2_{H_u}+m^2_{H_d}
\nonumber\\
&&m^2_{h_0}=\frac{1}{2}(m^2_{A^0}+m_Z^2-\sqrt{(m^2_{A^0}-m_z^2)^2+4m_Z^2m_{A^0}^2sin^2(2\beta)})
\nonumber\\
&&m^2_{H^0}=\frac{1}{2}(m^2_{A^0}+m_Z^2+\sqrt{(m^2_{A^0}-m_z^2)^2+4m_Z^2m_{A^0}^2sin^2(2\beta)})
\nonumber\\
&&m_{H^{\pm}}^2=m_W^2+m_{A^0}^2
\label{rel6565}
\end{eqnarray}

Here the relations $m_Z^2=\frac{g^2+g'^2}{2}(v_u^2+v_d^2)=\frac{g^2+g'^2}{2}v^2$  and $\tan{\beta}=\frac{v_u}{v_d}$ were employed.
From this by simple Taylor expansion one determines a higher bound on the lightest scalar:
\begin{eqnarray}
m_{h^0}<m_Z|\cos(2\beta)|.
\label{lim787}
\end{eqnarray}
\section{An exotic MSSM}

Inspired by the latest experimental discovery  of a light Higgs particle with the mass in the range $m_h=125-126$ GeV we propose
the following version with far-reaching consequences of the MSSM Lagrangian.
We change the sign of one of the chiral multiplets with the additional gauge-chiral multiplet interactions associated to it. We will discuss the case when this multiplet is the $H_d$ multiplet as the other case can be easily obtained from this. According to the analysis made in Chapter I this change will not affect the supersymmetric nature of the lagrangian as we have shown that the cancelations under the supersymmetry transformation are separate for the gauge and chiral supermultiplets and are also separate for the two chiral supermultiplets by
themselves. We also know that the superpotential should be invariant by itself so we consider exactly the same superpotential as in Eq(\ref{superp0990}). The main change will be in
the relative signs of the F terms (between them) and D terms (between them). We are mainly interested in the Higgs scalar sector. The contribution of  corresponding F terms to
the Lagrangian will be:
\begin{eqnarray}
{\cal L}_{auxiliar}=+F^{u*}F^u-F^{d*}F^d+W_uF^u+W_dF^d+....
\label{contr656}
\end{eqnarray}

This leads to the following term in the scalar potential:
\begin{eqnarray}
V=|\mu|^2(|H^0_u|^2+|H^+_u|^2)-|\mu|^2(|H^0_d|^2+|H^-_d|^2)+....
\label{contr5454}
\end{eqnarray}

Note that the $|\mu|^2$ terms have now opposite signs. In case we change the signs of the $H_d$ multiplet this will appear with the opposite sign.

The D terms appear in the Lagrangian:
\begin{eqnarray}
{\cal L}_D=+\frac{1}{2}D^aD^a+g_a(H_u^{\dagger}T^aH_u)D^a-g_a(H_d^{\dagger}T^aH_d)D^a
\label{d5454}
\end{eqnarray}

which leads to the following contribution in the scalar potential:
\begin{eqnarray}
V=\frac{1}{2}g_a^2(H_u^{\dagger}T^aH_u-H_d^{\dagger}T^aH_d)(H_u^{\dagger}T^aH_u-H_d^{\dagger}T^aH_d)
\label{pot76}
\end{eqnarray}
Here $T^a$ is a generic generator of the groups $SU(2)$ and $U(1)$. Then Eq. (\ref{pot76}) leads to:

\begin{eqnarray}
V=\frac{g^2+g'^2}{8}(|H^0_u|^2+|H^+_u|^2+|H^0_d|^2+|H_d^-|^2)^2-
\frac{1}{2}g^2|H_u^+H^{0*}_d+H^0_uH^{-*}_d|^2
\label{finres4343}
\end{eqnarray}

Finally we can write the complete potential obtained through this procedure and work with it.
\begin{eqnarray}
&&V=(|\mu|^2+m_{H_u}^2)(|H_u^0|^2+|H_u^+|^2)+(-|\mu|^2+m_{H_d}^2)(|H_d^0|^2+|H_d^+|^2)+b(H_u^+H_d^--H_u^0H_d^0)+c.c.+
\nonumber\\
&&+\frac{1}{8}(g^2+g'^2)(|H_u^0|^2+|H_u^+|^2+|H_d^0|^2+|H_d^-|^2)^2-
\frac{1}{2}g^2|H_u^+H_d^{0*}+H_u^0H_d^{-*}|^2.
\label{fullpot6565}
\end{eqnarray}

\section{The kinetic and mass scalar terms}

Here we will analyze in detail the changes that intervene in the scalar sector.
We start with the minimum equation which now will have the form:
\begin{eqnarray}
&&m_{H_u}^2+|\mu|^2-b\frac{v_d}{v_u}+\frac{g^2+g'^2}{4}(v_d^2+v_u^2)=0
\nonumber\\
&&m_{H_d}^2-|\mu|^2-b\frac{v_u}{v_d}+\frac{g^2+g'^2}{4}(v_u^2+v_d^2)=0
\label{minnew5454}
\end{eqnarray}

The scalar mass matrix  becomes:
\begin{eqnarray}
M^2_{h,H}=\left[
\begin{array}{cc}
b\frac{v_d}{v_u}+\frac{g^2+g'^2}{2}v_u^2&-b+\frac{g^2+g'^2}{2}v_uv_d\\
-b+\frac{g^2+g'^2}{2}v_uv_d&b\frac{v_u}{v_d}+\frac{g^2+g'^2}{2}v_d^2
\end{array}
\right].
\label{newm65657}
\end{eqnarray}

The mass matrix for the pseudoscalars remains unchanged whereas the mass matrix for the charged scalars changes to:

\begin{eqnarray}
M^2_{H_u^+,H_d^-}=
\left[
\begin{array}{cc}
b\frac{v_d}{v_u}-\frac{g^2}{2}v_d^2&b-\frac{g^2}{2}v_uv_d\\
b-\frac{g^2}{2}v_uv_d&b\frac{v_u}{v_d}-\frac{g^2}{2}v_u^2
\end{array}
\right].
\label{ch65995}
\end{eqnarray}

The biggest difference comes however from the gauge sector where now the mass of the Z boson will be given by:
\begin{eqnarray}
m_Z^2=\frac{g^2+g'^2}{2}(v_u^2-v_d^2)
\label{newexp656}
\end{eqnarray}

This suggest the following definitions:
\begin{eqnarray}
&&v_u=v\cosh(\beta)
\nonumber\\
&&v_d=v\sinh(\beta)
\nonumber\\
&&\tanh(\beta)=\frac{v_d}{v_a}
\label{res44343}
\end{eqnarray}

Note the difference with the standard MSSM: $\cot(\beta)$ is replaced by $\tanh(\beta)$. All other couplings will change accordingly.

We will show here in detail how a simultaneously diagonalization of the kinetic terms and potential terms can be obtained for the neutral scalars.
From the mass matrix in Eq. (\ref{ch65995}), Eq. (\ref{newm65657}) and Eq. (\ref{ch65995})one can compute the naive masses:

\begin{eqnarray}
&&m^2_{G^0}=m^2_{G^{\pm}}=0
\nonumber\\
&&m_{A^0}^2=2b\coth(2\beta)
\nonumber\\
&&m_{H^{\pm}}^2=m_{A^0}^2-m_W^2\cosh(2\beta)
\nonumber\\
&&m_{h,H}^2=m_{A^0}^2
\nonumber\\
&&m_{h,H}^2=m_Z^2\cosh(2\beta).
\label{ms35454}
\end{eqnarray}

However we will not consider these the final masses as the kinetic terms have not yet been diagonalized.
In the neutral sector the two rotation matrices that realize the diagonalization of the potential are:
\begin{eqnarray}
\left[
\begin{array}{c}
{\rm Re} H_u\\
{\rm Re} H_d
\end{array}
\right]
=
\left[
\begin{array}{cc}
\cos{\alpha}&\sin{\alpha}\\
-\sin{\alpha}&\cos{\alpha}
\end{array}
\right]
\left[
\begin{array}{c}
h\\
H
\end{array}
\right],
\label{sc3434}
\end{eqnarray}

for the neutral scalars and,
\begin{eqnarray}
\left[
\begin{array}{c}
{\rm Im }H_u\\
{\rm Im}H_d
\end{array}
\right]
=
\left[
\begin{array}{cc}
\cos{\gamma}&\sin{\gamma}\\
-\sin{\gamma}&\cos{\gamma}
\end{array}
\right]
\left[
\begin{array}{c}
G^0\\
A^0
\end{array}
\right],
\label{ps756}
\end{eqnarray}

for the neutral pseudoscalars.
It can be shown that for the mass matrices at hand the following relations hold:
\begin{eqnarray}
&&\sin{2\alpha}=-\tanh{\beta}
\nonumber\\
&&\sin{2\alpha}=-\sin{2\gamma}.
\label{ang6565}
\end{eqnarray}

We will choose $\beta$ positive, $\alpha$ negative and $\gamma$ also positive.

In order to diagonalize also the kinetic terms we need to consider a full $4\times4$ matrix as the vacuum in this case breaks CP symmetry. We write first:

\begin{eqnarray}
\left[
\begin{array}{c}
{\rm  Re}H_u\\
{\rm  Re}H_d\\
{\rm Im} H_u\\
{\rm Im}H_d
\end{array}
\right]
=
\left[
\begin{array}{cccc}
\cos{\alpha}&\sin{\alpha}&0&0\\
-\sin{\alpha}&\cos{\alpha}&0&0\\
0&0&\cos{\gamma}&\sin{\gamma}\\
0&0&-\sin{\gamma}&\cos{\gamma}
\end{array}
\right]
\left[
\begin{array}{c}
h\\
H\\
G^0\\
A^0
\end{array}
\right]=R\left[
\begin{array}{c}
h\\
H\\
G^0\\
A^0
\end{array}
\right]
\label{firstexp8787}
\end{eqnarray}

We now use a trick to further diagonalize the kinetic term. We multiply the $4\times4$ matrix in Eq (\ref{firstexp8787}) by another matrix K to the right such that to maintain the diagonalization  of the mass matrices. Then the full rotation matrix will be:
\begin{eqnarray}
RK=R\left[
\begin{array}{cccc}
a_1&0&0&0\\
0&0&0&a_2\\
0&0&a_3&0\\
0&a_4&0&0
\end{array}
\right]=
\left[
\begin{array}{cccc}
a_1\cos{\alpha}&0&0&a_2\sin{\alpha}\\
-a_1\sin{\alpha}&0&0&a_2\cos{\alpha}\\
0&a_4\sin{\gamma}&a_3\cos{\gamma}&0\\
0&a_4\cos{\gamma}&-a_3\sin{\gamma}&0
\end{array}
\right].
\label{full6767}
\end{eqnarray}

Here $a_1$, $a_2$, $a_3$ and $a_4$ are complex numbers which will be used for a proper normalization. Then we need to diagonalize
\begin{eqnarray}
\partial_{\mu}({\rm Re}H_u+i{\rm Im}H_u)^{\dagger}\partial^{\mu}({\rm Re}H_u+i{\rm Im}H_u)-
\partial_{\mu}({\rm Re}H_d+i{\rm Im}H_d)^{\dagger}\partial^{\mu}({\rm Re}H_d+i{\rm Im}H_d)
\label{rt5455}
\end{eqnarray}

in a context where the real part can become imaginary and the opposite since they are considered just regular fields.
Then the following ansatz realizes the proper diagonalization and normalization:
\begin{eqnarray}
&&a_1=a_3=(\cosh{2\beta})^{1/2}
\nonumber\\
&&a_2=a_4=i(\cosh{2\beta})^{1/2}
\label{diag5454}
\end{eqnarray}

Here we used the fact that $\alpha=-\gamma$. Then the proper kinetic term will become:
\begin{eqnarray}
(\cos^2(\alpha)-\sin^2(\alpha))[\partial_{\mu}h\partial_{\mu}h-\partial_{\mu}H\partial_{\mu}H+\partial_{\mu}G^0\partial_{\mu}G^0-\partial_{\mu}A^0\partial_{\mu}A^0].
\label{kinterm33}
\end{eqnarray}

Next we need to diagonalize the charged scalar mass matrix and kinetic term. We write:
\begin{eqnarray}
\left[
\begin{array}{c}
H_u^+\\
H_d^{-*}=
\end{array}
\right]
=R_{\beta^+}
\left[
\begin{array}{c}
G^+\\
H^+
\end{array}
\right]
=
\left[
\begin{array}{cc}
\sin{\beta_{+}}&\cos{\beta_{+}}\\
-\cos{\beta_{+}}&\sin{\beta_{+}}
\end{array}
\right]
\left[
\begin{array}{c}
G^+\\
H^+
\end{array}
\right].
\label{ort554}
\end{eqnarray}

We translate into $4\times4$ matrix for the real and imaginary parts of the fields:
\begin{eqnarray}
\left[
\begin{array}{c}
{\rm Re} H_u^+\\
{\rm Re}H_d^{-*}\\
{\rm Im}H_u^+\\
{\rm Im}H_d^{-*}
\end{array}
\right]=R_{\beta_{+}}
\left[
\begin{array}{c}
{\rm Re}G^+\\
{\rm Re}H^+\\
{\rm Im}G^+\\
{\rm Im}H^+
\end{array}
\right]=
\left[
\begin{array}{cccc}
\sin{\beta_{+}}&\cos{\beta_{+}}&0&0\\
-\cos{\beta_{+}}&\sin{\beta_{+}}&0&0\\
0&0&\sin{\beta_{+}}&\cos{\beta_{+}}\\
0&0&-\cos{\beta_{+}}&\sin{\beta_{+}}
\end{array}
\right]
\left[
\begin{array}{c}
{\rm Re}G^+\\
{\rm Re}H^+\\
{\rm Im}G^+\\
{\rm Im}H^+
\end{array}
\right].
\label{stform656}
\end{eqnarray}

Now we apply the same procedure as in the neutral scalar case. We multiply to the right by a complex matrix to obtain:
\begin{eqnarray}
\left[
\begin{array}{c}
{\rm Re} H_u^+\\
{\rm Re}H_d^{-*}\\
{\rm Im}H_u^+\\
{\rm Im}H_d^{-*}
\end{array}
\right]=R_{\beta_{+}}
\left[
\begin{array}{cccc}
b_1&0&0&0\\
0&0&0&b_2\\
0&0&b_3&0\\
0&b_4&0&0
\end{array}
\right]
\left[
\begin{array}{c}
{\rm Re}G^+\\
{\rm Re}H^+\\
{\rm Im}G^+\\
{\rm Im}H^+
\end{array}
\right].
\label{eq3232}
\end{eqnarray}

We take in order to diagonalize the kinetic terms:
\begin{eqnarray}
&&b_1=(\cosh{2\beta})^{1/2}
\nonumber\\
&&b_2=-i(\cosh{2\beta})^{1/2}
\nonumber\\
&&b_3=-(\cosh{2\beta})^{1/2}
\nonumber\\
&&b_4=i(\cosh{2\beta})^{1/2}
\label{fact5454}
\end{eqnarray}

Again the normalization of the states introduces a factor of $\cosh{2\beta}$ in the masses. The kinetic term after this procedure  will become:
\begin{eqnarray}
(\cos^2(\alpha)-\sin^2(\alpha))[\partial_{\mu}H^{+}\partial^{\mu}H^{+*}-\partial_{\mu}G^{+}\partial^{\mu}G^{+*}].
\label{finalkin4454}
\end{eqnarray}

Since the  Goldstone bosons  become states of negative energy we will need to ensure that they still have the correct properties such that to provide the longitudinal states fo
the $W^{\pm}$ bosons. We postpone this for section VII where we will show how to deal with such states quantum mechanically.
The correct and final masses are then:
\begin{eqnarray}
&&m^2_{G^0}=m^2_{G^{\pm}}=0
\nonumber\\
&&m_{A^0}^2=-2b\coth(2\beta)\cosh(2\beta)
\nonumber\\
&&m_{H^{\pm}}^2=2b\coth(2\beta)\cosh(2\beta)-m_W^2(\cosh(2\beta))^2
\nonumber\\
&&m_{h,H}^2=m_{A^0}^2
\nonumber\\
&&m_{h,H}^2=m_Z^2(\cosh(2\beta))^2.
\label{ms35454}
\end{eqnarray}

Note as a check of consistency that the negative states have negative masses whereas the regular states have positive masses.
\section{Neutralinos and charginos}

The mass term for the neutralinos in the Lagrangian is exactly as in the MSSM:
\begin{eqnarray}
{\cal L}=-\frac{1}{2}(\Psi^0)^TM_{N}\Psi^0+c.c
\label{nn887}
\end{eqnarray}

Assume that the wino and bino masses are very large such that the mixing between the Higgsion and gaugino can be neglected. Then
one needs to diagonalize only the lower matrix corresponding to the Higgssino.
\begin{eqnarray}
\left[
\begin{array}{cc}
{\tilde H}_d^0&{\tilde H}_u^0
\end{array}
\right]
\left[
\begin{array}{cc}
0&\mu\\
-\mu&0
\end{array}
\right]
\left[
\begin{array}{c}
{\tilde H}_d^0\\
{\tilde H}_u^0
\end{array}
\right]
\label{res54545}
\end{eqnarray}

The transformation matrix that diagonalizes the mass term is:
\begin{eqnarray}
\left[
\begin{array}{c}
{\tilde H}_d^0\\
{\tilde H}_u^0
\end{array}
\right]
=\frac{1}{\sqrt{2}}
\left[
\begin{array}{cc}
1&-1\\
1&1
\end{array}
\right]
\left[
\begin{array}{c}
{\tilde N}_1\\
{\tilde N}_2
\end{array}
\right]
\label{res3343}
\end{eqnarray}

Thus the mixing angle is just $\beta_N=\frac{\pi}{4}$ and the eigenvalues are simply $\mu$ and $-\mu$.
We apply the same procedure as in the previous case to the $4\times4$ matrix formed by the real and imaginary parts of the fermion fields to obtain for the kinetic term exactly zero.
This is mainly due to the precise value of the mixing angle. In the general case when one considers corrections to the mixing angle if these corrections are small they would translate into
a large normalization factor multiplying  the masses. Thus they can be considerably larger than $|\mu|$.

The charginos sector diagonalization can proceed similarly to that of the charged scalars. Since the topic is more intricate will be discussed in more detail in further work.

\section{States of negative energies}

We start with a simple case of two scalar fields, one with a positive kinetic energy the other one with negative kinetic energy which interact with each other. The corresponding Lagrangian(\cite{Trodden}) is:
\begin{eqnarray}
{\cal L}=\frac{1}{2}\partial_{\mu}\Phi_1\partial^{\mu}\Phi_1-\frac{1}{2}\partial_{\mu}\partial^{\mu}\Phi_2
-\frac{1}{2}m_1^2\Phi_1^2+\frac{1}{2}m_2^2\Phi_2-\lambda\Phi_1^2\Phi_2^2
\label{ex3232}
\end{eqnarray}

The equations of motion in the absence of the interaction term are those of regular Klein-Gordon fields for both states:
\begin{eqnarray}
&&(\partial_{\mu}\partial^{\mu}+m_1^2)\Phi_1=0
\nonumber\\
&&(\partial_{\mu}\partial^{\mu}+m_2^2)\Phi_2=0
\label{motion7676}
\end{eqnarray}

They are modified in the presence of the interaction term to:
\begin{eqnarray}
&&(\partial_{\mu}\partial^{\mu}+m_1^2)\Phi_1+2\lambda)\Phi_1\Phi_2^2=0
\nonumber\\
&&(\partial_{\mu}\partial^{\mu}+m_2^2)\Phi_2-2\lambda\Phi_2\Phi_1^2=0
\label{somth66}
\end{eqnarray}

The counterpart of the minimum state in the regular case is a saddle point at $\langle\Phi_1\rangle=0$ (local minimum) and $\langle\Phi_1\rangle=0$ (local maximum). This can be seen from:
\begin{eqnarray}
&&\frac{\partial V}{\partial \Phi_1}=m_1^2\Phi_1+2\lambda\Phi_1\Phi_2^2=0
\nonumber\\
&&\frac{\partial V}{\partial \Phi_1}=-m_2^2\Phi_2+2\lambda\Phi_2\Phi_1^2=0
\nonumber\\
&&\frac{\partial^2 V}{\partial\Phi_1^2}|_0=m_1^2
\nonumber\\
&&\frac{\partial^2 V}{\partial\Phi_2^2}|_0=m_2^2
\nonumber\\
&&\frac{\partial^2 V}{\partial \Phi_1 \partial_2 \Phi_2}|_0=0
\label{min656}
\end{eqnarray}

This kind of model has been studied at the classical level in \cite{Trodden} where it has been shown that it can display stable oscillation for some range of parameter in particular if
$\lambda(\frac{M^2}{2m_1^2})<1$ for M the initial displacement of the oscillators.
However we still have a system with violates some of the principles on which we based our description of reality.

We are mostly interested however in the quantum behavior of this kind of system. One might consider for that two approaches described below.

\subsection{States with negative energies}

We consider for simplicity the Lagrangian of Eq (\ref{ex3232}):
\begin{eqnarray}
{\cal L}=\frac{1}{2}\partial_{\mu}\Phi_1\partial^{\mu}\Phi_1-\frac{1}{2}\partial_{\mu}\partial^{\mu}\Phi_2
-\frac{1}{2}m_1^2\Phi_1^2+\frac{1}{2}m_2^2\Phi_2
\label{newlagr878}
\end{eqnarray}

We then quantize both fields in exactly the same way such that:
\begin{eqnarray}
&&\Phi_i(\vec{x})=\int \frac{d^3 p}{2\pi^3}\sqrt{\frac{1}{2 E_i}}(a_{p_i}e^{i\vec{p_i}\vec{x}}+a_{p_i}^{\dagger}e^{-i\vec{p_i}\vec{x}})
\nonumber\\
&&\Pi_i(\vec{x})=\int \frac{d^3 p}{2\pi^3}(-i)\sqrt{\frac{E_i}{2}}(a_{p_i}e^{i\vec{p_i}\vec{x}}-a_{p_i}^{\dagger}e^{-i\vec{p_i}\vec{x}}).
\label{res43443}
\end{eqnarray}

Here the energies are positive,
\begin{eqnarray}
&&E_1=\sqrt{p_1^2+m_1^2}
\nonumber\\
&&E_2=\sqrt{p_2^2+m_2^2}
\label{rel878}
\end{eqnarray}

and the following commutation relations hold:
\begin{eqnarray}
&&[a_{p_i},a_{p_i'}^{\dagger}]=(2\pi)^3\delta^{(3)}(\vec{p}_i-\vec{p'}_i)
\nonumber\\
&&[a_{p_i},a_{p_j}]=[a_{p_i},a_{p_j}^{\dagger}]=[a^{\dagger}_{p_i},a^{\dagger}_{p_j}]=0\,\,\,for\,i\neq j
\label{com76776}
\end{eqnarray}

Then the hamiltonian is:
\begin{eqnarray}
H=\int d^3 p \{E_{1}[a_{p_1}^{\dagger}a_{p_1}+\frac{1}{2}[a_{p_1},a_{p_1}^{\dagger}]]-E_{2}[a_{p_2}^{\dagger}a_{p_2}+\frac{1}{2}[a_{p_2},a_{p_2}^{\dagger}]]\}
\label{res443}
\end{eqnarray}

and the operator $a_{p_1}^{\dagger}$ creates a state of positive energy $E_{1}$ whereas the operator $a_{p_2}^{\dagger}$ creates a state of negative energy $-E_{2}$.
One can goes further to study interactions of this type. The main difference will be that in the conservation of momentum-energy the incoming states with negative energy will behave like
outgoing states with positive energy and the opposite. When dealing with decays or scatterings (see \cite{trodden}) one sees that infinities already appear at tree level as the phase space
is infinite. Furthermore unwanted cascade decays might appear which make the appearance of negative states also impossible from the experimental point of view.
There is also another problem:what happens when we are dealing with charged states? It turns out that this question together with the detailed diagonalization treatment from section V gives the correct answer to how one should deal with these states quantum mechanically.

\subsection{Exotic states}

First note that negative energy is equivalent with time reversal and this, if CPT conservation holds, should be equivalent with states with reversed CP and positive energies.
States with negative energies appeared from the dawn of quantum field theory when it was noticed that the Dirac equation has two solutions one with positive energy the other one with negative energy. The problem was later solved by quantizing the theory such that the states of negative energy correspond to the antiparticles of positive energies.
To show that let us consider the hamiltonian of a Dirac particle with the quantum expression:
\begin{eqnarray}
H=\int d^3 x\Psi^{\dagger}{\cal H}\Psi=\sum_k\hbar\omega_k a_k^{\dagger}a_k.
\label{dir54646}
\end{eqnarray}

This in general is not a positive operator as there are always solutions with both $\omega_k$ positive and negative. The appearance of the unwanted negative energy states can be cured
if one introduces the antiparticle operators :
\begin{eqnarray}
&&b_k^{\dagger}=a_k
\nonumber\\
&&b_k=a_k^{\dagger}
\label{def43535}
\end{eqnarray}

Thus the hamiltonian is extended to include both the particle and the antiparticle,
\begin{eqnarray}
H=\sum_k \hbar\omega_k a_k^{\dagger}a_k+\sum_k \hbar\omega_k b_k^{\dagger}b_k+E_0
\label{newh554}
\end{eqnarray}
and it is a sum of energies of two types of oscillators, corresponding to the electron and positron but both with positive energies.

We now implement this procedure to our case:
For a simple Klein-Gordon field charged or not we simply replace the operators in Eq. (\ref{res43443}) by:
\begin{eqnarray}
&&\Phi_2=\int \frac{d^3p}{(2\pi)^3}\sqrt{\frac{1}{2\omega_{p_2}}}[-b_{p_2}^{\dagger}e^{i\vec{p}_2\vec{x}}+b_{p_2}e^{-i\vec{p}_2\vec{x}}]
\nonumber\\
&&\Pi_2=\int \frac{d^3p}{(2\pi)^3}(-i)\sqrt{\frac{\omega_{p_2}}{2}}[-b_{p_2}^{\dagger}e^{i\vec{p}_2\vec{x}}-b_{p_2}e^{-i\vec{p}_2\vec{x}}].
\label{new43443}
\end{eqnarray}

Note that again the correct commutation relations hold:
\begin{eqnarray}
[b_p,b_{p'}^{\dagger}]=-[b_p^{\dagger},b_p]=(2\pi)^3\delta^{(3)}(\vec{p}-\vec{p'})
\label{com6767}
\end{eqnarray}

Then the contribution of negative states to the hamiltonian becomes:
\begin{eqnarray}
H_2=-\omega_2[a_{p_2}^{\dagger}a_{p_2}+\frac{1}{2}[a_{p_2}a_{p_2}^{\dagger}]]=
\omega_2[b_{p_2}^{\dagger}b_{p_2}+\frac{1}{2}[b_{p_2}b_{p_2}^{\dagger}]]
\label{finres5554}
\end{eqnarray}

Thus the energy becomes again positive definite. Note that this procedure is equivalent to simple replacement of the particle by its CP conjugate.
The usual rules of energy momentum conservation apply  when we deal with decays and interactions.
However some consistency checks are in order to be sure that this approach works properly. All interactions that include couplings of the negative states
with their hermitian conjugate are safe. The potential problems might appear from the following type of terms:
\begin{eqnarray}
&&\mu H_u H_d
\nonumber\\
&&b(H^0_uH^0_d-H_u^{+}H_d^{-})
\label{prob66657}
\end{eqnarray}

Here a replacement of the particle by its CP conjugate can lead at first glance to disaster. However after finding the mass eigenstates
and their couplings one observes that quite general a negative state does not couple anymore with the normal state but with its CP conjugate such that in the end
one obtains the correct invariant couplings (see for details our diagonalization procedure).

Another relevant aspect that we need to take into account refers to the possible gauge anomalies and to the correct contribution of the Goldstone bosons to
the longitudinal states of the $W^{\pm}$ in particular.
First we need to check that the up and down Higgsino chosen with opposite Y as in the standard MSSM still lead to the cancelation of the gauge anomalies.
In our model the couplings of the down Higgsino with the gauge fields  have an opposite sign with respect to the corresponding coupling of the up Higgsino
so they will contribute to the anomaly with factor of $(-1)^3=-1$. On the other hand for each diagram the fermion is replaced by its CP conjugate such that the
contribution of down Higgsino reduces (for example for the $U(1)^3$ anomaly) to:

\begin{eqnarray}
{\rm Tr}[Y^3]={\rm Tr}[Y_u]^3-{\rm Tr}[-Y_d^3]=0.
\label{asnom}
\end{eqnarray}

Note that only the correct quantization of the negative states leads to the cancelation of the anomalies. Also by construction this does not affect
the couplings of the Higgs bosons with the down quarks.

In order to show that our negative states charged Goldstone bosons still account for the longitudinal states of the charged W bosons we need to write explicitly  the
gauge eigenstates in terms of the mass eigenstates in this sector:
\begin{eqnarray}
&&H_u^{+}=\sin(\beta_+)G^{+*}-\cos(\beta_+)H^+
\nonumber\\
&&H_d^{-}=-\cos(\beta)G^{+}+\sin(\beta)H^{+*}
\label{diag665}
\end{eqnarray}

The relevant interaction vertex is given by the term in the Lagrangian:
\begin{eqnarray}
-ig\partial_{\mu}H_u^{+*}v_u(W_1^{\mu}-i W_2^{\mu})
-ig\partial_{\mu}H_d^- v_d(W_1^{\mu}-i W_2^{\mu})
\label{vert5454}
\end{eqnarray}
which leads to,

\begin{eqnarray}
-ig\partial_{\mu}G^{\mu+}\sin(\beta_+)v_u-ig\partial_{\mu}G^{\mu+}\cos(\beta_+)v_d
\label{res554}
\end{eqnarray}

Thus the contribution to the polarization amplitude turns to be exactly that of the regular MSSM:
\begin{eqnarray}
i\frac{g^2}{2}v^2g^{\mu\nu}-i\frac{g^2}{2}\frac{k^{\mu}k^{\nu}}{k^2}
\label{p00}
\end{eqnarray}

 The correct diagonalization of the charged states was crucial in obtaining this result.

Quite general our model can be easily treated as a regular theory with positive energy states with the added feature that the overall masses and couplings are simpler and closer
to the desired ones from the experimental point of view.

\section{Discussion}

We first use Eq (\ref{ms35454}) and the experimental value of $125.9$ GeV for the mass of the Higgs boson to determine all the angles in the model:
\begin{eqnarray}
&&\beta=0.424
\nonumber\\
&&\alpha=-0.380
\nonumber\\
&&\gamma=0.380
\label{angles54554}
\end{eqnarray}

The couplings of the Higgs boson with the $W^{\pm}$, Z, t and b quarks are exact those of the standard model.
This can be seen from:
\begin{eqnarray}
&&(h \bar{t}t)=(\cosh(2\beta))^{1/2}\frac{\cos(\alpha)}{\cosh(\beta)}(h\bar{t}t)_{SM}=(h\bar{t}t)_{SM}
\nonumber\\
&&( h \bar{b} b)=(\cosh(2\beta))^{1/2}\frac{-\sin(\alpha)}{\sinh(\beta)}(h\bar{t}t)_{SM}=(h\bar{b}b)_{SM}
\nonumber\\
&&(h W_{\mu}^{+}W^{-\mu})=(\cosh(2\beta))^{1/2}(\cos(\alpha)\cosh(\beta)+\sin(\alpha)\sinh(\beta))(h W_{\mu}^{+}W^{-\mu})_{SM}=(h W_{\mu}^{+}W^{-\mu})_{SM}
\label{res443}
\end{eqnarray}

Apparently the two photon decay rate in this model if one considers only the top and charged W loops is exactly that of the  standard model as the couplings are the same.
Although charged scalars may appear in the loop their estimated large mass leads to a very small contribution. An increase of the diphoton rate of the Higgs boson can be explained in this
model only by the contributions of the superpartners in the loops.

In conclusion our exotic MSSM contains all the particles and interactions of the MSSM with some major differences.
First of all the suggested mass of $125.9$ GeV of the light Higgs boson can be obtained at tree level. All the mixings and angles in the neutral scalars sector are predicted and are very simple. Then the main couplings of the Higgs with the top, bottom, Z and W bosons are exactly those of the standard model thus making impossible the distinction between the two models in the absence of the superpartners. The masses of the neutral Higgsino can attain very large values for small mixings due to the normalization factors.

Our purpose here was to present a new theoretical model and give some hints about the phenomenological implications. Other phenomenological aspects
will be discussed in detail elsewhere.

\section*{Acknowledgments} \vskip -.5cm
I am happy to thank J. Schechter for support and encouragement and for useful comments on the manuscript.
This work has been supported by PN 09370102/2009.

\end{document}